\documentclass[manuscript,acmsmall]{acmart}

\usepackage{subfigure}
\usepackage{graphicx}
\usepackage{listings}
\usepackage{multirow}
\usepackage[ruled,vlined,linesnumbered]{algorithm2e}

\SetCommentSty{mycommfont}

\definecolor{amber}{rgb}{1.0, 0.49, 0.0}

\newcommand{\sidong}[1]{\textcolor{red}{\textbf{Sidong}: #1}}

\newcommand{\specialcell}[2][c]{%
  \begin{tabular}[#1]{@{}c@{}}#2\end{tabular}}

\newcommand{\tool}{\texttt{Auto-Icon+}\xspace} 

\AtBeginDocument{%
  \providecommand\BibTeX{{%
    \normalfont B\kern-0.5em{\scshape i\kern-0.25em b}\kern-0.8em\TeX}}}

\setcopyright{acmcopyright}
\acmJournal{TIIS}
\acmYear{2022} \acmVolume{1} \acmNumber{1} \acmArticle{1} \acmMonth{1} \acmPrice{15.00}\acmDOI{10.1145/3531065}

\begin{document}

\title{\tool: An Automated End-to-End Code Generation Tool for Icon Designs in UI Development}

\author{Sidong Feng}
\email{sidong.feng@monash.edu}
\affiliation{%
	\department{Faculty of Information Technology}
	\institution{Monash University}
	\city{Melbourne}
	\country{Australia}
}

\author{Minmin Jiang}
\affiliation{%
	\institution{Alibaba Group}
	\city{Hangzhou}
	\country{China}}
\email{minchao.jmm@alibaba-inc.com}

\author{Tingting Zhou}
\affiliation{%
	\institution{Alibaba Group}
	\city{Hangzhou}
	\country{China}}
\email{miaojing@taobao.com}

\author{Yankun Zhen}
\affiliation{%
	\institution{Alibaba Group}
	\city{Hangzhou}
	\country{China}}
\email{zhenyankun.zyk@alibaba-inc.com}

\author{Chunyang Chen}
\affiliation{%
	\department{Faculty of Information Technology}
	\institution{Monash University}
	\city{Melbourne}
	\country{Australia}}
\email{chunyang.chen@monash.edu}

\renewcommand{\shortauthors}{Feng et al.}

\begin{abstract}
  Approximately 50\% of development resources are devoted to UI development tasks~\cite{beaudouin2004designing}.
Occupying a large proportion of development resources, developing icons can be a time-consuming task, because developers need to consider not only effective implementation methods but also easy-to-understand descriptions.
  In this paper, we present \tool, an approach for automatically generating readable and efficient code for icons from design artifacts.
  According to our interviews to understand the gap between designers (icons are assembled from multiple components) and developers (icons as single images), we apply a heuristic clustering algorithm to compose the components into an icon image.
  We then propose an approach based on a deep learning model and computer vision methods to convert the composed icon image to fonts with descriptive labels, thereby reducing the laborious manual effort for developers and facilitating UI development.
  We quantitatively evaluate the quality of our method in the real world UI development environment and demonstrate that our method offers developers accurate, efficient, readable, and usable code for icon designs, in terms of saving 65.2\% implementing time.
\end{abstract}

\ccsdesc[500]{Human-centered computing~Empirical studies in accessibility}

\keywords{code accessibility, icon implementation, neural networks}

\maketitle

\section{Introduction}
A user interface (UI) consists of series of elements, such as text, colors, images, widgets, etc.
Designers are constantly focusing on icons as they are highly functional in a user interface~\cite{koutsourelakis2010icons,black2017icons,islam2015exploring,malik2017sustainable}.
One of the biggest benefits of icons is that they can be universal. 
For instance, by adding a red “X” icon to your user interface design, users are informed that clicking this icon leads to the closure of a component.
Furthermore, icons can make UIs look more engaging. For example, instead of using basic bullets or drop-downs filled with words, a themed group of icons can capture instant attention from users.
Consequently, icons become an elegant yet efficient way to communicate with and help guide user through experience. 

Despite of all these benefits, icons have three fundamental limitations in the day-to-day development environment, in terms of \textit{transition gap}, \textit{rendering speed} and \textit{code accessibility}.
First, it is a challenging task for developers to implement icons from the design artifacts as many designers follow their own design styles which are highly differ from each other.
For example, some of them design an icon with many small components, though just one combined image in the icon implementation for optimizing the network traffic and caching.
Such gap adds the complexity and effort for developers to preprocess the design draft before icon implementation.
Second, to ensure a smooth user interaction, UI should be rendered in under 16ms~\cite{gomez2016mining,slowrendering,slowrendering2}, while icon implemented as an image faces the slow rendering problem, due to image download speed, image loading efficiency, etc.
These issues will directly affect the quality of the product and user experience, requiring more effort from developers to develop an advanced method to overcome the problem.
Third, in the process of UI implementation, many developers directly import the icon resources from the design artifacts without considering the meaning of the content, resulting in poor description/comment during coding.
Different codes render the same visual effect to users, while it is different for developers to develop and maintain.
A non-descriptive code increases the complexity and effort required to developers as they need to look at the associated location of the UI to understand the meaning of the code.

This challenge motivates us to develop a proactive tool to address the existing UI development limitations and improve the efficiency and accessibility of code.
Our tool, \tool involves four main features.
First, to bridge the conceptual gap between icon design and icon implementation, we propose a heuristic machine learning technique to automated agglomerate the scattered icon components into clusters.
The component clusters can then be composed to icon images, which can be applied in various downstream tasks related to improving the rendering efficiency (i.e., icon font, css clipping) and code accessibility (i.e., description, color).
Second, to meet the requirement of efficient rendering, we develop an automated technique to convert icon image to icon font, which is a typeface font.
Once the font is loaded, the icon will be rendered immediately without downloading the image resources, thereby reducing HTTP requests and improving the rendering speed. Icon font can further optimize the performance of rendering by adopting HTML5 offline storage.
Besides, icon font has other potential attributes that can facilitate UI development, such as easy to use (i.e., use the CSS's @fontface attribute to load the font), flexible (i.e., capable to change color, lossless scale), etc.
Third, understanding the meaning of icons is a challenging problem.
There are numerous types of icons in the UIs. Icons representing the same meaning can have different styles and can be presented in different scales as shown in Table~\ref{tab:categorization}. Also, icons are often not co-located with texts explaining their meaning, making it difficult to understand from the context. 
In order to offer an easy access for developers to develop through understanding the meaning of icons, we collect 100k icons from existing icon sharing website Alibaba Iconfont~\cite{iconfont} - each associating with a label described by designer.
By analyzing the icons and labels, we construct 100 categories, such as "left", "pay", "calendar", "house", etc. 
We then train a deep learning classification model to predict the category of the icon as its description.
The experiments demonstrate that our model with the average accuracy as 0.87 in an efficient classification speed as 17.48ms, outperforms the other deep learning based models and computer vision based methods.
Third, to provide more accessibility to developers on the description of icon images, we also detect the primary color of icons by adopting HSV color space~\cite{sural2002segmentation}.
We refer to our mechanism tool \tool to build an intelligent support for developers in the real context of UI development, assisting developing standardized and efficient code.

To demonstrate the usefulness of \tool, we carry out an user study to show if our tool for automatically converting an icon to a font with label descriptions can help provide more knowledge on code accessibility and accelerate UI development for developers.
After analyzing ten professional developers' feedback with all positive responses on our mechanism tool and we find that the code for icon generated by our tool can achieve better readability compared with the code manually written by professional developers.
Besides, \tool has been implemented and deployed in Alibaba \textit{Imgcook} platform. The results demonstrates that our tool provides 84\% usable code for icon designs in a realistic development situations.
Our contributions can be summarized below:
\begin{itemize}
    \item We identify the fundamental limitations of existing UI development of icon images.
    The informal interviews with professional developers also confirm these issues qualitatively.
    \item To bridge the gap between icon design and icon implementation, we develop a machine-learning based technique to compose the components of icon in the design artifact.
    \item Based on the emerging 100 icon categories, we develop deep-learning and computer-vision based techniques for specifically converting icon to font with label describing its meaning and color to provide developers understand knowledge of code.
    \item We conduct large-scale experiments to evaluate the performance of our tool \tool and shows that our tool achieves good accuracy compared with baselines.
    The evaluation conducted with developers and tested on the real-world development platform demonstrates the usefulness of our tool.
    \item We contribute to the community by offering intelligent support for developers to efficiently implement icon designs comply with code standardization.
\end{itemize}

\section{Related Works}

\subsection{Transition Gap}
UI design and implementation require different mindset and expertise. 
The former is performed by user experience designers and architects via design tools (e.g., Sketch~\cite{sketch}, Photoshop~\cite{ps}), while the latter performed by developers via development tools (e.g., Android Studio~\cite{androidstudio}, Xcode~\cite{xcode}). 
Existing researches well support these two phases respectively~\cite{lee2020guicomp, yang2021uis, chen2021should, todi2021conversations, bunian2021vins, zhang2021screen, ang2021learning,feng2021auto}.
For example, Li et al.~\cite{li2021screen2vec} provide a holistic representation of UI to help designers quickly build up a realistic understanding of the design space for an app feature and get design inspirations from existing UI designs.
But few of them support effective transition from UI design artifacts to UI implementation.

Supporting this transition is challenging due to the gap between designers and developers and the complexity of design artifacts (see Fig~\ref{fig:iconexample}).
Some researches lower the transition gap by translating UI screenshots into UI implementation~\cite{chen2018ui,moran2018machine,sketch2codemicrosoft,sketch2codeairbnb, chen2019storydroid,wu2021screen}.
Nguyn et al.~\cite{nguyen2015reverse} follow the mobile specific heuristics and adopt hybrid method based on Optical Character Recognition (OCR) and image processing to generate a static code from UI screenshots.
Pix2Code~\cite{beltramelli2018pix2code} adopts an iterative encoder/decoder deep learning model consisting of a CNN for extracting UI screenshots features, and a LSTM decoder for generating the UI code tokens.
However, these works are unclear if the UI implementations are realistic or useful from a developer’s point of view, as the implementation of the approaches are only validated on a small set of synthetically UI screenshots.
It is difficult to judge how well the approaches would perform on real UI screenshots. 
In contrast, \tool is tailored for the design artifact driven development practice, and is aimed at solving real-world UI development challenges, such as developers requiring manual effort to compose the icon from scattered components in the design artifact.
There has also been both commercial and academic work related to design artifact driven development for creating high-fidelity design artifacts~\cite{fluidui, mockup, proto}, deploying real-device prototype environments~\cite{meskens2009plug,meskens2009shortening}, and verifying design violations~\cite{moran2018automated}.
However, such tools and approaches tend to either impose too many restrictions on designers or do not allow for direct creation of code, thus the icon composition problem still persists in practice.
To address this, we adopt a machine-learning approach to automatically agglomerate the scattered icon components into clusters through heuristic correlation coefficient to bridge the gap between icon design and icon implementation.



\subsection{UI Rendering}
Ensuring fast rendering speed is an essential part in UI development, since slow rendering creates poor user experience.
Many studies focus on improving rendering speed via reducing bugs~\cite{carette2017investigating,li2019characterizing,nejati2016depth,huang2017shuffledog,rosen2017push}.
In contrast, we focus on analyzing image displaying performance in UI rendering.
There are a few related works in this domain.
For example, Systrace~\cite{systrace} is a tool that allows developers to collect precise timing information about UI rendering on devices. 
However, it does not provide any suggestions for improvement.
To address this problem, many studies introduce reliable approaches to improve rendering efficiency such as image resizing based on pattern-based analysis~\cite{liu2014characterizing}, a manual image resource management based on resource leakage analysis~\cite{wu2016light}.
Gao et al.~\cite{gao2017every} implement a system called DRAW which aims to reveal UI performance problems in an application such as excessive overdraw and slow image components detection. 
With the suggestion of the image displaying performance analysis by DRAW, developers can manually improve the rendering performance of slow image displaying.
While these works offer image management suggestions to developers to achieve better rendering performance, they still need to be improved manually.
In contrast, we propose an image conversion technology based on computer vision and graphic algorithms to convert icons into font types for achieving faster UI rendering.

\subsection{Code Accessibility}
Digital devices such as computer, mobile phone and tablets are widely used.
To ensure the quality of software, many research works have been conducted~\cite{butler2010android, fu2013people, liu2022guided}.
Most of these works focus on the functionality and usability of apps such as GUI design~\cite{chen2019gallery,feng2022gallery, chen2019storydroid,chen2020object,xie2020uied,liu2020owl,zhao2021guigan}, GUI animation linting~\cite{zhao2020seenomaly,zhao2019actionnet}, localization~\cite{wang2019domain,feng2021gifdroid,xie2020uied}, privacy and security~\cite{chen2019gui,chen2018automated,dehling2015exploring,feng2019mobidroid}, and performance~\cite{linares2015developers,zhao2016novel}.
Few research works are related to accessibility issues.
Some works in Human-Computer Interaction area have explored the accessibility issues of mobile apps~\cite{ yan2019current,wang2012measurement,mora2017comprehensive,king2016government,chen2020unblind}.
In these work, the lack of description in image-based components in UI is commonly regarded as an important accessibility issue.
For instance, Harrison et al.~\cite{harrison2011kineticons} establish an initial ‘kineticon vocabulary’ containing a set of 39 kinetic behaviors for icon images, such as spin, bounce, running, etc.
Ross et al.~\cite{ross2018examining} identify some common labeling issues in Android apps via analyzing the icon image labeling.
With crowd source method, Zhang et al~\cite{zhang2018robust} annotate GUI elements without content description.
However, these works still require support from developers.
Due to the increasingly developed Convolutional Neural Networks (CNNs) technologies, dramatic advances appears in the field of image classification which is applied to automatically annotate tags for images.
Chen et al.~\cite{chen2020lost} analyze the tags associated with the whole GUI artwork collected from Dribbble, and emerge an vocabulary that summarizes the relationship between the tags. 
Based on the vocabulary, they adopt a classification model to recommend the general tags in the GUI, such as "sport", "food", etc.
Different from their work, we predict more fine-grained categories, such as "football", "burger", etc.
And also, they focus on predicting the categories of the whole UI which is subjective to human perception, but the categories of small icons are usually more intuitive.
A similar work to ours is the icon sensitive classification by Xiao et al~\cite{xiao2019iconintent}.
They utilize traditional computer vision techniques like SIFT and FAST to extract the features of icons and classify icons into 8 categories through calculating their similarity.
After the systematically investigation of icons, we discover the fundamental limitations in icons discussed in Section~\ref{empiricalstudy}, in terms of \textit{high cross-class similarity} and \textit{small, transparent and low contrast}. These findings conflict with methods applied in their paper such as applying rotation to augment the dataset. Moreover, we show that deep learning model is fruitful for the icon classification problem than the tradition computer vision technique in Section~\ref{prediction_result}.
In our work, according to the characteristic of icons, we propose a deep learning model to automatically classify icons in a more fine-grained (100) category and also adopt a computer vision technique to detect its primary color.
\section{Preliminary Study}
\label{pre_study}
To better understand the challenges in the real-world UI development environment, we conducted an interview with 12 front-end developers from the big companies.
Two authors first developed the interview protocol, and conducted pilot studies with two participants. 
Based on the pilot studies, we refined the interview protocol and conducted 10 interviews formally.
The average length of these interviews is 20 minutes.
We started with general questions, such as questions about working years, workload of development, and number of projects developed.
Then, we asked the interviewees how they developed the code for icon images.
We particularly asked what motivated them to adopt the approach, whether they revised the implementation, what approaches could achieve the same effect, what they perceived as the impact of the implementation, how the implementation behaved in the process of development and is there any difference on UI development between personal projects and company tasks.

\begin{figure*}  
	\centering 
	\includegraphics[width=0.65\linewidth]{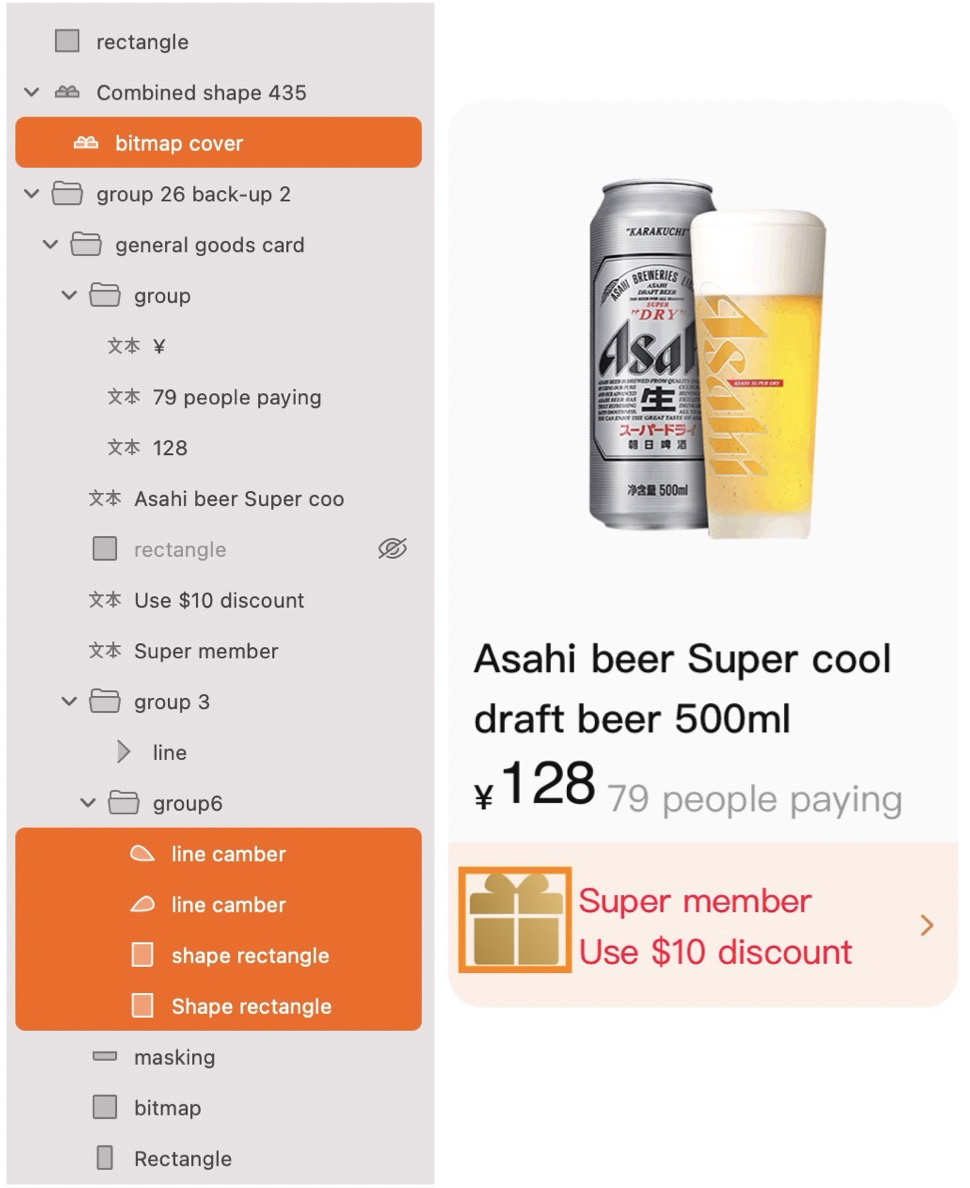}  
	\caption{An illustration of icon design in the design artifact given by designers. A "gift" icon is composed by multiple components such as bitmap, lines, and shapes (highlighted).}
	\label{fig:iconexample}
\end{figure*}

\subsection{Research Question 1: do developers implement icons directly from the design artifacts?}
\label{em_rq1}

Once the design artifacts are completed by designers, they are handed off to development teams who are responsible for implementing the designs in code. 
Fig.~\ref{fig:iconexample} shows an example of the design artifact handed off to developers.
The design artifacts are stored as archives containing JSON encoded layout and a number of binary assets such as bitmap images, curves, shapes~\cite{sketch_developer}.
The layout is structured as a cumulative hierarchy comprising a tree structure, starting with a single root node, where the spatial layout of parent always encompasses contained child components.
A discrete component with a corresponding set of attributes can be represented as a five-tuple in the form ($<name>, <x-position,y-position>, <width,height>, <text>, <image>$). 
Here the first element of the tuple describes the name of the component.
The next four elements of the tuple describe the location of the top left point for the bounding box of the component, and the height and width attributes describe the size of the bounding box.
The text attribute corresponds to text displayed (if any) by the component and the image attribute represents an image (if any) of the component with bounds adhering to the position attributes.

All of the developers demonstrated that they always preprocess the icon designs before implementing, due to the gap to designers.
As D9 said that \textit{"We are all fulfilling of our goals, but in different directions."}
At the core of it, designers are focused on the graphical/visual representations of icons, while the developers take care of the functionality of icons, such as efficient rendering.
However, this gap points to a more prominent problem for icon implementation in practice as D10 showcased an icon design in the design artifact (as shown in Fig~\ref{fig:iconexample}) and said:

\textit{
In most big IT or software companies, the visual designers design the UI with professional tools like Sketch~\cite{sketch} or Adobe Photoshop~\cite{ps}.
These tools provide build-in curvature pen/pencil tool allow designers to create some complicated components by drawing smooth curves and straight line segments with equal ease in order to sketch out the desired shape.
Well-designed icons are usually composed of images, curves, lines, shapes, etc.
For example, in Fig~\ref{fig:iconexample}, the "gift" icon is composed by lines, shapes, and bitmaps.
Developers need to spend extra effort to compose all the small pieces into an icon to gain icon rendering efficiency and code accessibility. 
}


They also mentioned that finding all the components of an icon is surprisingly non-trivial task.
This is because that as designers only concern with the visual representation of icon, they may create one icon with many components scattered all over the design artifact. 
For example, the bitmap ("knot-bow") is a component of the "gift" icon but not in the same hierarchy of other components in Fig~\ref{fig:iconexample}.
D9 supported this claim that:

\textit{
Although there is much information (i.e., name, attribute, hierarchical relationship, etc.) in the design artifact, that information does not align well with that required in the icon composition and implementation.
In practice, I potentially consider three characteristics to compose an icon.
First, the rendering of the components of icon should be close/overlap to each other.
Second, some designers may organize the resources of an icon with some layout hierarchies to group them, for example "group6" in Fig~\ref{fig:iconexample}.
Third, some designers may use the same format to name icon components, for example "gift-piece-x".
Therefore, it takes time to search all the components with reference to their rendering, hierarchy, and attribute.
}

\noindent\fbox{
    \parbox{0.98\textwidth}{
        Due to the gap between designers and developers, developers can hardly implement icons from design artifacts directly.
        Given a design artifact, developers require to manually compose the icon image by searching all the related components based on their attributes, hierarchies and renderings, adding to the complexity and effort required to icon implementation in practice.
    }
}

\subsection{Research Question 2: do developers implement icons in font or images?}
By summarising the approaches, we collected 4 ways of rendering icons, i.e., image tag <img> or <svg>, icon tag <i>, css background image, and custom tag <SvgIcon> as shown in Table~\ref{tab:icon_development}.
One third of our developers listed all approaches, 80\% developers knew the way of using image and font.
There are 2 developers who have never heard of or used the fonts to render icons, D2 said:

\textit{Making front-end development is fun, although sometimes it hurts because I do not have adequate learning experience.
There are few front-end courses in universities, and these courses usually contain relatively simple knowledge, such as what is <div> block, how to connect HTML and CSS together, etc.
They do not teach the usage of font, especially they do not distinguish the difference between fonts and images in rendering icons.}

70\% developers implemented icons as image when developing front-end codes based on UI design draft files because they found that converting icon to font is a complicated and laborious process. For example, D7 mentioned:

\textit{To implement the approach of icon font, I first need to upload the image to the existing conversion websites such as icomoon~\cite{icomoon} and Fontello~\cite{fontello}.
Then, I need to download the generated icon font to my local device.
Last but not least, I need to copy the generated CSS code to CSS files.
This entire process requires a lot of time and effort, but due to time constraints, the process is not compatible in industry.}

One developer D2 from Alibaba described how limit the time in their UI development:

\textit{
Every year, Alibaba has more than 240 events which stores offer special discount, such as Double 11 Global Delight Event, Tmall Thanksgiving Day Event, 1212 Global Discount Event, etc.
Due to the high demand for the UI development in the duration of events, we are required to implement UIs in 3 or 4 days.}

Developers also considered the trade-off between UI performance and its value. 
Since the usage of icon font does not provide business value, it is often in a low priority in industry. 
Even if they knew the benefit of using font, they would not put effort in doing this. 
For instance, D9 explained:

\textit{Although I know the icon font is better compared to icon image, I will not apply this approach in development.
I usually have 3 tasks in a week, such as UI implementation, bug testing, algorithm implementation, etc.
I agree that icon font can improve UI rendering performance and provide better user experience.
But, the overall functionality will still work without icon font. 
In contrast, without bug testing, the front-end codes may not work, resulting in significantly impact on the company business.  
And if I do not implement the algorithm, other developers will not be able to apply the API in their development, which will slow down the development speed and delay the product release time.
}

Another example shows the potential gap between industry and individual is that 50\% developers mentioned that they use font to render icons when developing their personal projects, such as homepage, blog, tutorial, etc. 
For example, D2 said,

\textit{When I developed my first personal website, I discovered Font Awesome~\cite{fontawesome}, a font toolkit to render icons by simply adding class description.
Since this is my website, I can design freely according to my preferences.
To quickly develop my website, I used the font in Font Awesome to implement all the icons in my website.
However, it is not applicable in industry.
In industry, every icon is well designed according to the company culture and design specifications. Therefore, it is not suitable to apply widely used icon font resources from online platforms.
In addition, using online icon fonts involves intellectual property (IP) issues which must be avoided in the industry.}

\noindent\fbox{
    \parbox{0.98\textwidth}{
        Despite most of the developers know the benefits of using font to render icons, few of them implement font in practice.
        The icon they used is distinct to the online resources as it comprises company culture and design guidelines. 
        Therefore, rather than directly using the online resources, developers have to spend extra effort in converting icons to font, which is time-consuming and laborious.
    }
}


\subsection{Research Question 3: do developers write descriptions for icons?}
\label{rq2}
All of them mentioned that they did write descriptions/comments in their personal projects, such as assignments, homepage, etc.
However, half of developers did not write descriptions in practice due to the following practical reasons.
First, since the readability of code is not a mandatory requirement, many developers did not write well-formatted descriptions for code. 
For example, the code in the industry cannot be released as open-source.
As D9 said that \textit{"Since our code can not be released to public, I would not spend too much time on writing comments in code because only a few internal developers would collaborate on my tasks."}
In addition, since updating iteration in the industry is fast, it is not worth to put too much effort in commenting, especially for icons.
For example, D10 said, 
\textit{In the year of 2019, our company developed over 1 million UIs.
Due to the diversity of UIs, few designs are re-implemented and few code are reviewed. Because of the fast updating iteration and low reusing rate, I did not write well-formatted comment, particularly for the images.
I was developing a shopping application which images cover more than half of the UIs.
To develop the large amount of images quickly, I prefer using <img> tag without any alternative description.
}

Second, 80\% developers mentioned that writing a well understood description is a challenging task. 
It requires developers to understand the intention of the icons, while few developers pay attention to the content of the UIs.
For example, D7 explained,

\textit{I agree that the clear descriptions in the code can keep the code readable and “save lives”, while unreasonable descriptions “kill lives”. 
However, it is hard to write a good description. 
Here is the process of how I write the descriptions: 
Firstly, I design a comment for every component, image, ..., based on its characteristic. 
Secondly, I rename and simplify the comments according to practical requirement. 
Thirdly, I check if the comments match the content of UIs or not. 
Then I repeat this process until the deadline.
And obviously, the process is time-consuming and not applicable in the industry.}

Despite the insufficient descriptions in the code may not impede professional developers, it creates a significant cognitive burden for interns and new developers.
For example, D3 said,

\textit{I am a junior student who came to the company for internship.
The first task assigned to me by my leader was to understand the code.
However, I found that most of the code is uncommented, which makes it very difficult for me to understand.
To understand this part of code, I asked more than 5 developers who participated this project.
These uncommented codes negatively influenced my work. 
}

\noindent\fbox{
    \parbox{0.98\textwidth}{
        Developers rarely write descriptions for images, especially for icons, because the loose restriction on code readability makes developers less cared about code descriptions.
        Most of developers agree with difficulty on designing simple, concise and easy-understood  descriptions.
        The lack of description can adversely affect novice employees and lead to inadequate understanding of the code.
    }
}

\section{Icon Characteristics}
\label{characteristics}


\begin{table}
\caption{The 40 icon classes identified through an iterative open coding of 100k icons from the Iconfont~\cite{iconfont}.}
\centering
\small
\begin{tabular}{p{0.12\textwidth}p{0.45\textwidth}p{0.2\textwidth}c}
    \toprule
    \bf{CLASS} & \bf{ASSOCIATED LABEL} & \bf{EXAMPLES} & \bf{NUMBER} \\
    \midrule
    \bf{\small add} & \small{plus, addition, increase, expand, create} & \parbox[c]{1em}{\includegraphics[width=1in]{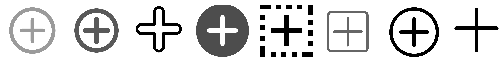}} & 357 \\
    \bf{\small calendar} & \small{date, event, time, planning} & \parbox[c]{1em}{\includegraphics[width=1in]{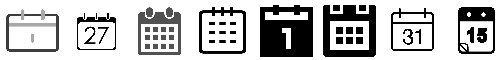}} & 324 \\
    \bf{\small camera} & \small{photo, take-photo} & \parbox[c]{1em}{\includegraphics[width=1in]{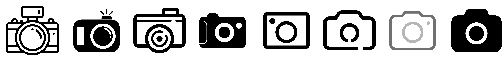}} & 355 \\
    \bf{\small chat} & \small{chat-bubble, message, request, comment} & \parbox[c]{1em}{\includegraphics[width=1in]{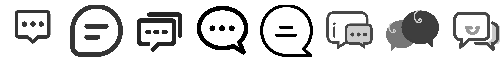}} & 372 \\
    \bf{\small complete} & \small{finish, confirm, tick, check, ok, done} & \parbox[c]{1em}{\includegraphics[width=1in]{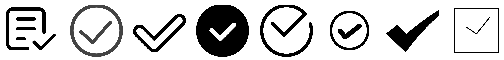}} & 432 \\
    \bf{\small computer} & \small{laptop, device, computer-response, desktop} & \parbox[c]{1em}{\includegraphics[width=1in]{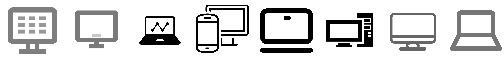}} & 521 \\
    \bf{\small crop} & \small{prune, crop-tool, shear, clipper crop-portrait} & \parbox[c]{1em}{\includegraphics[width=1in]{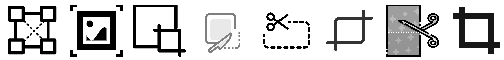}} & 436 \\
    \bf{\small download} & \small{file-download, save, import, cloud} & \parbox[c]{1em}{\includegraphics[width=1in]{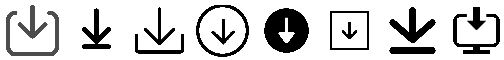}} & 444 \\
    \bf{\small edit} & \small{editing, handwriting, pencil, pen, edit-fill, modify} & \parbox[c]{1em}{\includegraphics[width=1in]{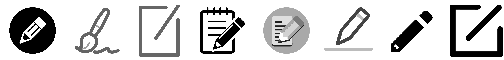}} & 546 \\
    \bf{\small emoji} & \small{amojee, sad, happy, emotion} & \parbox[c]{1em}{\includegraphics[width=1in]{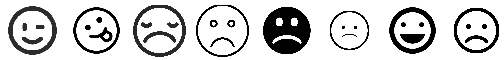}} & 374 \\
    \bf{\small envelope} & \small{letter, email, mail, inbox} & \parbox[c]{1em}{\includegraphics[width=1in]{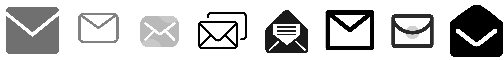}} & 332 \\
    \bf{\small exit} & \small{quit, close, switch-off, logout} & \parbox[c]{1em}{\includegraphics[width=1in]{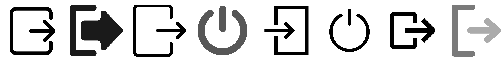}} & 404 \\
    \bf{\small flower} & \small{flowers, flower pot, sunflower, valentine-flower} & \parbox[c]{1em}{\includegraphics[width=1in]{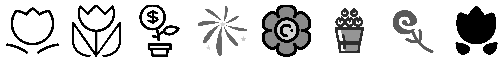}} & 377 \\
    \bf{\small gift} & \small{present, reward, surprise} & \parbox[c]{1em}{\includegraphics[width=1in]{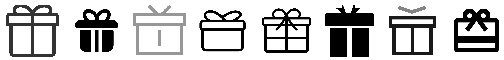}} & 340 \\
    \bf{\small house} & \small{home, rent, house-area, house asset, building, mall} & \parbox[c]{1em}{\includegraphics[width=1in]{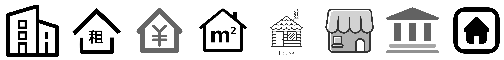}} & 378 \\
    \bf{\small left} & \small{return, back, prev, backwards} & \parbox[c]{1em}{\includegraphics[width=1in]{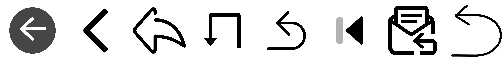}} & 531 \\
    \bf{\small like} & \small{thumb-up, heart, vote, hand-like, upvote, dislike, favourite} & \parbox[c]{1em}{\includegraphics[width=1in]{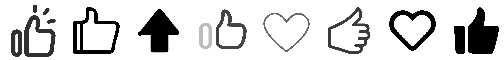}} & 386 \\
    \bf{\small location} & \small{gps, direction, compass, navigation} & \parbox[c]{1em}{\includegraphics[width=1in]{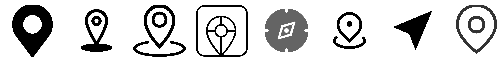}} & 543 \\
    \bf{\small menu} & \small{menu file, card, menufold, menu-line, more, dashboard} & \parbox[c]{1em}{\includegraphics[width=1in]{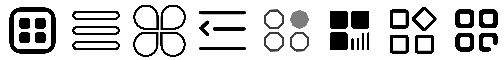}} & 351 \\
    \bf{\small minus} & \small{remove, minus (with circle), minus-sign} & \parbox[c]{1em}{\includegraphics[width=1in]{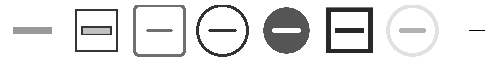}} & 556 \\
    \bf{\small music} & \small{music-note, music-library, musical-instrument} & \parbox[c]{1em}{\includegraphics[width=1in]{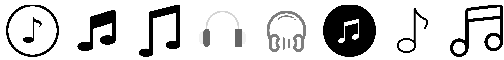}} & 375 \\
    \bf{\small news} & \small{newspaper, info, announcement} & \parbox[c]{1em}{\includegraphics[width=1in]{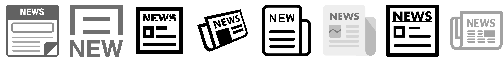}} & 423 \\
    \bf{\small package} & \small{package-up, package-sent, handpackage, personal package} & \parbox[c]{1em}{\includegraphics[width=1in]{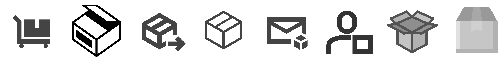}} & 362 \\
    \bf{\small pay} & \small{money, wallet, dollar, commerce} & \parbox[c]{1em}{\includegraphics[width=1in]{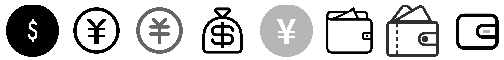}} & 364 \\
    \bf{\small person} & \small{user, avatar, account, customer} & \parbox[c]{1em}{\includegraphics[width=1in]{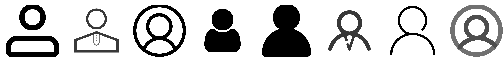}} & 562 \\
    \bf{\small photo} & \small{image, picture, camera} & \parbox[c]{1em}{\includegraphics[width=1in]{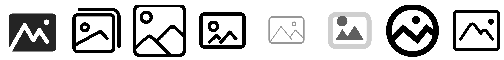}} & 481 \\
    \bf{\small play} & \small{playicon, broadcast, play voice, play button, play arrow} & \parbox[c]{1em}{\includegraphics[width=1in]{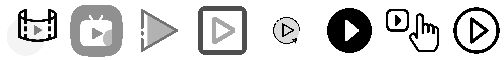}} & 498 \\
    \bf{\small question} & \small{ask, faq, information, help, info, support} & \parbox[c]{1em}{\includegraphics[width=1in]{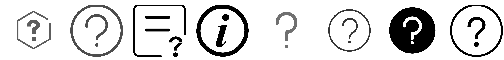}} & 350 \\
    \bf{\small refresh} & \small{reload, sync, reset, recreate} & \parbox[c]{1em}{\includegraphics[width=1in]{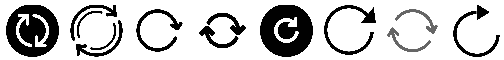}} & 321 \\
    \bf{\small right} & \small{forward, next, go, arrow-forward} & \parbox[c]{1em}{\includegraphics[width=1in]{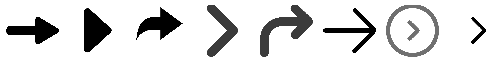}} & 412 \\
    \bf{\small safe} & \small{safe box, safety, safety certificate, lock, secure} & \parbox[c]{1em}{\includegraphics[width=1in]{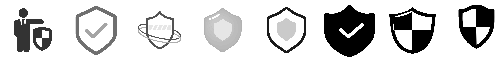}} & 476 \\
    \bf{\small search} & \small{investigate, search-engine, magnifier, find, glass} & \parbox[c]{1em}{\includegraphics[width=1in]{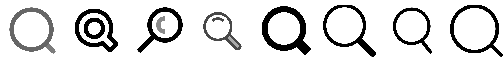}} & 377 \\
    \bf{\small send} & \small{send-arrow, paper-plane, message} & \parbox[c]{1em}{\includegraphics[width=1in]{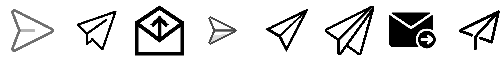}} & 318 \\
    \bf{\small settings} & \small{toolbox, gear, preferences, options} & \parbox[c]{1em}{\includegraphics[width=1in]{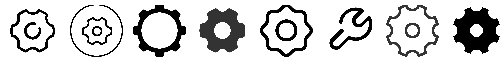}} & 317 \\
    \bf{\small shopping} & \small{cart, shopping-bag, checkout} & \parbox[c]{1em}{\includegraphics[width=1in]{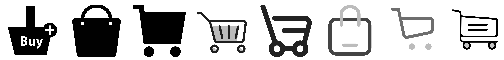}} & 472 \\
    \bf{\small signal} & \small{signal-tower, wave, radio, broadcast} & \parbox[c]{1em}{\includegraphics[width=1in]{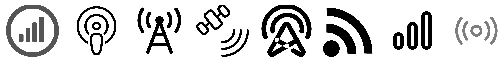}} & 548 \\
    \bf{\small sound} & \small{speaker, sound volume, player} & \parbox[c]{1em}{\includegraphics[width=1in]{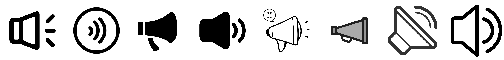}} & 415 \\
    \bf{\small star} & \small{collection, rate, favourite} & \parbox[c]{1em}{\includegraphics[width=1in]{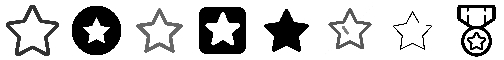}} & 424 \\
    \bf{\small switch} & \small{switch-on/off, switcher, open, close} & \parbox[c]{1em}{\includegraphics[width=1in]{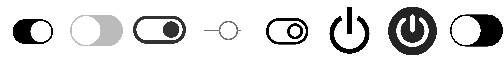}} & 319 \\
    \bf{\small text} & \small{word, textbox, font, size} & \parbox[c]{1em}{\includegraphics[width=1in]{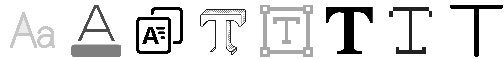}} & 446 \\
    \bf{\small visibility} & \small{visible, show, hide, visibility-off, in-sight} & \parbox[c]{1em}{\includegraphics[width=1in]{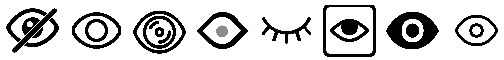}} & 339 \\
    \bf{\small warn} & \small{alarm, warning, error, report, alert} & \parbox[c]{1em}{\includegraphics[width=1in]{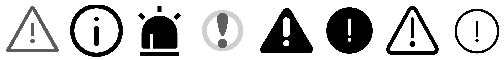}} & 369 \\
    \bf{\small wifi} & \small{wi-fi, wireless, network, signal} & \parbox[c]{1em}{\includegraphics[width=1in]{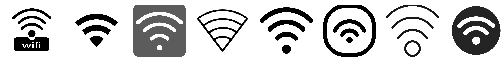}} & 429 \\
    \bf{\small zoom-in} & \small{fullscreen, expand, adjust, magnifier} & \parbox[c]{1em}{\includegraphics[width=1in]{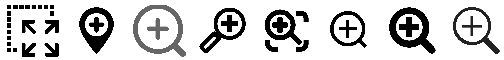}} & 384 \\
    \bottomrule
\end{tabular}
\label{tab:categorization}
\end{table}

In this section, we carry out an empirical study of collaborative icon labelling in online icon design sharing websites to understand its characteristics for motivating the required tool support.
There are numerous online icon design sharing websites such as Font Awesome~\cite{fontawesome}, Google Material Design Icons~\cite{googlematerial}, provide comprehensive icon library to assist designers and developers in designing and coding.
In these online icon websites, each label matches only one icon design.
While in real case, one label may have several different designs, revealing the limited diversity of these websites.
In this work, we select Alibaba Iconfont website~\cite{iconfont} as our study subject - not only because it has gained significant popularity among designers' community, but also due to it has became repositories of knowledge with millions of diverse icon designs created by designers.
To collect icons and associated labels, we built a web crawler based on the Breadth-First search strategy~\cite{najork2001breadth} i.e., collecting a queue of URLs from a seed list, and putting the queue as the seed in the second stage while iterating.
The crawling process continued from December 12, 2019 to July 1, 2020 with a collection of 100k graphical icons.

\subsection{Overview}
\label{empiricalstudy}

During the process of open coding the categories of icons semantic, we find that one label can be written in different styles. 
For example, the label "crop" can be written in not only its standard format, but also its derivations synonyms like "prune", "clip", "crop-tool".
Moreover, due to the icon labelling process in Iconfont is informal and icon designs are contributed by thousands of designers with very diverse technical and linguistic backgrounds the same concept may labeled in many user defined terms such as "crop-fill", "crop-portrait", "icon-crop-solid-24px".
The wide presence of forms poses a serious challenge to icon classification task.
For example, the icon can be described to the class of "crop" or "clip", which makes sense in both classes.

To address the problem, we adopted association rule mining~\cite{agrawal1994fast} to discover label correlations from label co-occurrences in icons.
We leveraged the visual information from the icons and textual information from the labels to group a pair-wise correlation of labels.
For measuring the visual similarity, we adopted the image similarity score MSE~\cite{wang2004image} $sim_{vis(x,y)}$ to calculate the likelihood if two icons are the same.
For measuring the textual information, we first trained a word embedding~\cite{mikolov2013distributed} model to convert each label into a vector that encodes its semantic.
Then we defined a lexical similarity threshold based on the string edit distance~\cite{levenshtein1966binary} $sim_{text(x,y)}$ to check if two labels are similar enough in the form.
The labels are grouped as a pair-wise correlation if $sim_{vis(x,y)}\geq0.9$ or $sim_{text(x,y)}\geq0.9$.
As we wanted to discover the semantics and construct a lexicon of categories, we found frequent pairs of labels.
A pair of labels is frequent if the percentage of how many icons are labelled with this pair of tags compared with all the icons are above the minimum support threshold $t_{sup}\geq0.001$.
Given a frequent pair of labels $\{t_1, t_2\}$, association rule mining generated an association rule $t_1 \Rightarrow t_2$ if the confidence of the rule $t_{conf}\geq0.2$.
Given the mined association rules, we constructed an undirected graph $G(V, E)$, where the node set $V$ contains the labels appearing in the association rules, and the edge set $E$ contains undirected edges $<t_1, t_2>$ (i.e., pair of label associations) if the two labels have the association $t_1 \Rightarrow t_2$ or $t_2 \Rightarrow t_1$.
Note that the graph is undirected because association rules indicate only the correlations between antecedent and consequent.
All threshold values were carefully selected through manually check, considering the balance between the information coverage and overload.

To identify the set of frequently occurred icon label categories, we performed an iterative open coding of most frequent co-occurring labels (or approximately 9.2\% of the dataset 542,334 in total) with existing expert lexicon of categories in books and websites such as \textit{Google’s Material icon set}~\cite{googlematerial}, \textit{IBM's Design Language of Iconography}~\cite{ibmicon} and \textit{Design Pattern Gallery}~\cite{neil2014mobile}.
Two researchers from our team independently coded the categories of these labels, noting any part of the initial vocabulary.
Note that both researchers have design experiences in both icons and UI development.
After the initial coding, the researchers met and discussed the discrepancies and the set of new label categories until consensus was reached.
A semantic icon categories can be seen in Table~\ref{tab:categorization}.
We observed two distinct characteristics in icons compared to the physical-world objects.

\textbf{High cross-class similarity:} 
Icons of different classes often have similar size, shape and visual features.
The visual differences to distinguish different classes of icons can be subtle, particularly small widgets are differentiated by small visual cues.
For example, the difference between "newspaper" and “file" lies in a text of news at the top/bottom side of "newspaper", while a plus/minus symbol distinguishes
"zoom\underline{ }in"/"zoom\underline{ }out" from "search".
In addition, direction is also an important aspect to distinguish classes. For example, the inclined waves represent "signal" and the upward waves represent "wifi".
Existing object classification tasks usually deal with physical objects with distinct features across classes, for example, fishes, flowers, hockey and people in the popular ImageNet dataset~\cite{deng2009imagenet}.
High cross-class similarity affects classification as the class can be not easily distinguished.

\textbf{Small, transparent and low contrast:} 
To make UI unique and stylish in the screen, icons are usually small and partially transparent, such as the last icon in the "minus" class shown in Table~\ref{tab:categorization}.
The transparent icons in the UIs do not cause vision conflict, while they are less visible when separated from the background context.
For example, the first icon in the "text" class in Table~\ref{tab:categorization} is an icon with low color contrast and uses transparency and shadow to stress contrast.
While the contrast of the object is obvious in the current dataset, especially apparent in the greyscale format such as MNIST dataset~\cite{lecun1998gradient}.

\noindent\fbox{
    \parbox{0.98\textwidth}{
        Existing icon sharing sites contain a wide presence of forms of labeling.
        Based on different background knowledge, designers use different same-meaning labels to annotate the same icon.
        Such limitation not only confirms our finding of difficulty of commenting in Section~\ref{rq2}, but also hinders the potential challenge in classification task.
        Therefore, a data mining approach capturing visual and textual information is applied to construct a lexicon in icons.
        By observing the lexicon, we find that two distinct characteristics of icon different from the existing physical object orientated dataset.
    }
}
\begin{figure*}  
	\centering 
	\includegraphics[width=\linewidth]{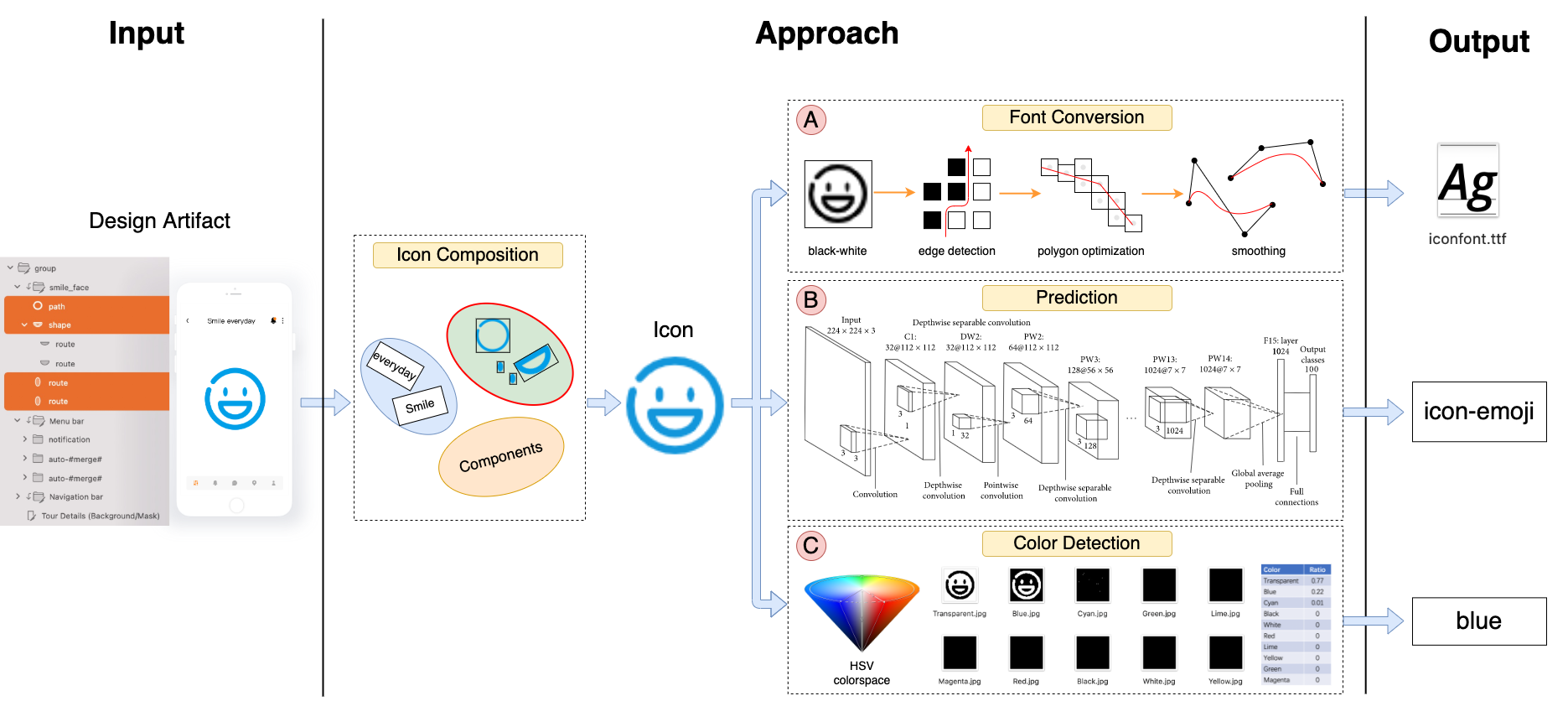}  
	\caption{The approach of our tool, \tool. Given a design artifact as input, we first compose the components into icons by clustering, for example, the "eye", "mouth", "face" components are composed into a smiling face icon. We then apply three major functions, involving font conversion, prediction and color detection, to generate the output code for the icon, which consists of a font file (.ttf) and a descriptive tag, for example "icon-emoji" and "blue".}
	\label{fig:approach}
\end{figure*}

\section{Approach}

Based on the interview study in Section~\ref{pre_study}, we summarized three design considerations for our approach: 
(1) icons are assembled from multiple components in the design artifact.
(2) icons implemented as fonts can expedite rendering.
(3) adding descriptions to icons can increase readability and accessibility of code.

To solve (1), we proposed a machine learning technique to cluster the scattered components in the design artifact through heuristic correlation coefficient and then composed the clusters of components into icons (Section~\ref{iconcomposition}).
To solve (2), we proposed an automated conversion technique, taking an icon as the input, and outputting a vector graphics font(Section~\ref{convertion}). 
To address (3), we got inspired from the findings in our study on icon characteristic in Section~\ref{characteristics} to develop a deep learning model to automatically assign the classes of icons in order to reduce the effort of manually designing the description of icons (Section~\ref{prediction}).
Additionally, we applied a primary color detection method based on computer vision to keep track of the primary color of the icon in order to support more detailed description in code (Section~\ref{detection}). 
The overview of our approach is shown in Fig.~\ref{fig:approach}.

\subsection{Icon Composition}
\label{iconcomposition}

\begin{algorithm}[t]
    \linespread{0.2}
	\SetAlgoLined
	\SetNoFillComment
	\SetKwInOut{Input}{Input}
	\SetKwInOut{Output}{Output}
	\Input{UI design artifact $X$ with $n$ elements $\{x_1, x_2, x_3, ..., x_n\}$}
	\Output{Hierarchical Dendrogram $HD$}
	construct $n \times n$ matrix $HD$ with correlation metric $cor(i,j)$ between the elements \;
	\While{len($X$) > 1}{
	    Select the pair $(x_i,x_j)$ with the largest correlation value, such as $x_i,x_j \in X$ \;
	    Merge the pair $(x_i,x_j)$ into a new cluster $x_{merge}=x_i \cup x_j$, let $x_i,x_j$ be the sub clusters of cluster $x_{merge}$. \;
	    Update $X \gets X \cup \{x_{merge}\} - \{x_i,x_j\}$ \;
	    \ForEach{$x \in X$}{
	        Update the matrix $HD$ with correlation metric $cor(x,x_{merge})$ \;
	    }
	}
	return $HD$
	\caption{Hierarchical Agglomerative Clustering (HAC)}
	\label{algorithm:hac}
\end{algorithm}

The most common way to group related pieces is through clustering.
Normally, the assumed number of clusters may be unreliable since the number of the icons among the design artifact is unknown and thus the top-down partitioning methods (i.e., K-Means~\cite{likas2003global}, Expectation-Maximization~\cite{dempster1977maximum}, etc.) will not applicable.
To provide clustering without requiring the knowledge of clusters, we adopted a bottom-up approach Hierarchical Agglomerative Clustering (HAC)~\cite{murtagh2014ward}.
The details of HAC is shown in Algorithm~\ref{algorithm:hac}, where each component in the design artifact is treated as a singleton cluster to start with and then they are successively merged into pairs of clusters until all components have merged into one single large cluster.
The main parameters in this algorithms are the metric used to compute the correlation value of components, which determines the pair of clusters to be merged at each step.
According to our observations in Section~\ref{em_rq1}, we defined a heuristic correlation metric that takes into account component's attribute, hierarchy, and rendering:
\begin{equation}
	Correlation(x,y) = \alpha \times ATTR(x,y) + \beta \times HRCHY(x,y) + \gamma \times IOU(x,y)
	\label{eq:correlation}
\end{equation}
where $ATTR(x,y)$ measures the attribute type of two components $x,y$, that is if they have same type (i.e., image, curve, shape, etc.), it assigns the value to 1 else 0.
$HRCHY(x,y)$ measures the hierarchy between components, the value is assigned to 1 if they are under the same group, 0 otherwise.
$IOU(x,y)$ measures the overlap between components, taking a value between 0 to 1.
And $\alpha, \beta, \gamma$ is the user-defined weights for each measurements.
The higher the correlation value, the more likely the cluster can composed to an icon.

\begin{figure*}  
	\centering 
	\includegraphics[width=0.75\linewidth]{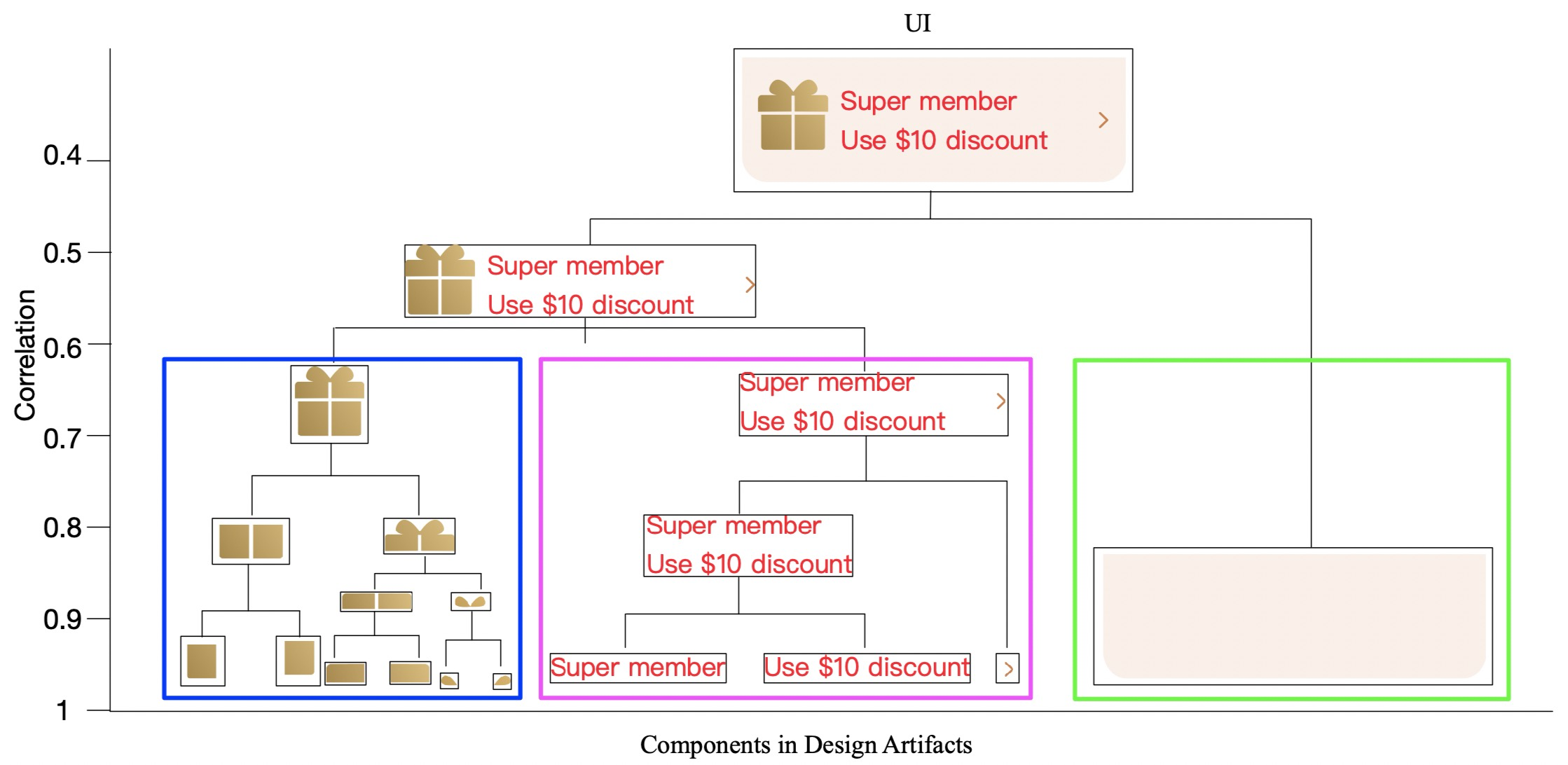}  
	\caption{The HAC clustering algorithm to compose components into icons in the design artifacts, that repeatedly selecting and merging pairs of clusters based on a heuristic correlation metric, until a single all-inclusive cluster (UI). By empirically setting a correlation threshold to 0.6 and distinguishing icon cluster by its feature as discussed in Section~\ref{empiricalstudy} (small, transparent and low contrast), the blue cluster is determined as an icon.}
	\label{fig:hac-example}
\end{figure*}

The agglomeration of clusters results in a tree-like structure called the dendrogram as shown in Fig~\ref{fig:hac-example}.
The value is highest at the lowest level of the dendrogram and it decreases as the clusters merge into the final single cluster. 
By cutting the dendrogram at an empirically setting of value threshold (e.g., 0.6), we may retrieved several clusters, for example, the icon (blue cluster), text (red cluster), and background (green cluster) in Fig~\ref{fig:hac-example}.
To further identify if the aggregated cluster is an icon (blue cluster), we followed the feature of icon discussed in Section~\ref{empiricalstudy}, that the icon should be small and low contrast.


\subsection{Font Conversion from Icon}
\label{convertion}
Unlike converting font to image, transcribing image to font, which is also known as image tracing problem, is a difficult task.
In this work, we adopted the state-of-the-art Potrace~\cite{selinger2003potrace} in Fig~\ref{fig:approach}A.
We first applied a pre-processing method for converting color to binary (i.e., black and white) image by setting a threshold to control the bit of each pixel after calculating the average value in three channels \((R + G + B) / 3 \).
We regarded the pixel is of white if the average value is larger than 128, while black if the value is equal to or smaller than 128.
Then, we detected the edge in the black-white image.
An edge is defined to be a border between a white pixel and a black pixel, which indicate which pixels from the original image constitute the borders of region.
Note that the edge is assigned a direction so that when moving from the first endpoint to the second, the black pixel is on the left (as shown in Fig~\ref{fig:approach}A edge detection).
This process was repeated until we reached the starting point, at which point we have found a closed path which encloses a black region.
Once the border was found, we approximated/optimized the border with a polygon to figure out which border pixels is possible to connect with a straight line such that the line passes through all the border pixels between its endpoints. 
To detect the optimal polygon, we computed a penalty value to measure the average distance from the edge to the pixels it approximates. 
The polygon with the smallest penalty (equivalent to the polygon with the fewest pixels) is the optimal one.
Finally, we used a cubic curve defined by four control points (also known as Bezier curve~\cite{sederberg1992approximation}) to smooth the corners.
The first and fourth control points (i.e., midpoints of the edges of the polygon) give the locations of the two endpoints of the curve, while the second and third (i.e., chosen on the polygon edges through the endpoints) indicate the direction and magnitude of the derivative of the curve at each endpoint.

\subsection{Prediction for Icon}
\label{prediction}
Traditional Convolutional Nerual Network (CNN)~\cite{lecun1998gradient, krizhevsky2012imagenet} has shown great potential as a solution for difficult vision problems.
MobileNetV2~\cite{sandler2018mobilenetv2} distills the best practices in convolutional network design into a simple architecture that can serve as competitive performance but keep low parameters and mathematical operations to reduce computational power.
The architecture of the network is shown in Fig~\ref{fig:approach}B.

Instead of using regular convolutional layers widely used in traditional CNN architectures to capture essential information from images but are expensive to compute, MobileNetV2 adopted a more advanced one, depthwise separable convolutions.
Depthwise separable convolution combined a $3*3$ convolution layer and two $1*1$ convolution layers.
The $1*1$ convolution layer (also named as pointwise convolution layer) was used to combine the filter values into new features, while the $3*3$ convolution (also called as depthwise convolution layer) was used to filter the input feature map.
Inspired from the dimension augmentation in the work of~\cite{lin2013network}, MobileNetV2 used a $1*1$ pointwise convolution layer to expand the number of channels in the input feature map.
Then it used a $3*3$ depthwise convolution layer to filter the input feature map and a $1*1$ convolution layer to reduce the number of channels of feature map.
The network borrowed the idea of residual connection in ResNet~\cite{he2016deep} to help with the flow of gradients.
In addition, batch normalization and activation layer were added between each depthwise convolution layer and pointwise convolution layer to make the network more stable during training.
For detailed implementation, we adopted the stride of 2 in the depthwise convolution layer to downsample the feature map.
For the first two activation layers, the network used ReLU6 defined as $y=min(max(0,x),6)$ because of its robustness in low-precision computation~\cite{howard2017mobilenets}, and a linear transformation (also known as Linear Bottleneck Layer) was applied to the last activation layer to prevent ReLU from destroying features.

\subsection{Color Detection of Icon}
\label{detection}
Since the conversion between icon and font sacrifices the color identity, we added an attribute to keep track of the primary color of the icon.
To that end, we adopted HSV colorspace for color detection.
We first removed the fourth alpha channel as transparent and made a conversion from RGB color to HSV colorspace. Each RGB color has a range of HSV vale. The lower range is the minimum shade of the color that will be detected, and the upper range is the maximum shade. 
For example, blue is in the range of $\langle 100,43,46 \rangle$--$\langle 124,255,255 \rangle$.
Then, we created a mask for each color (black, blue, cyan, green, lime, megenta, red, white) as shown in Fig.~\ref{fig:approach}C. 
The mask is the areas that HSV value on pixels match the color between the lower range and upper range. 
Finally, we calculated the area of the mask in each color and the corresponding image occupancy ratio. 
The color with the maximum ratio was identified as the primary color of the icon (the blue in the example in Fig.~\ref{fig:approach}C).

\section{Experiments}
In this section, we first set up an experiment to analyze the performance of our tool. Then we conduct a pilot user study to evaluate the usefulness of our tool. Furthermore, we demonstrate its usefulness on a large-scale industrial benchmark. 
The goal of our experiments is to answer the following research questions, in terms of accuracy, efficiency and applicability.
\textit{RQ 1: how accurate is our heuristic algorithm in composing icons from design artifacts?}
\textit{RQ 2: how accurate is our model in predicting labels for icon images?}
\textit{RQ 3: how much do our tool increase the efficiency of UI development?}
\textit{RQ 4: what are the developers' opinions on the usability of our tool?}

\subsection{Icon Composition}
\subsubsection{Dataset:}
To evaluate the performance of our heuristic composition algorithm, we compare our automatically generated icon to manually generated ground truth.
To generate the ground truth, we had one author and 10 non-author paid annotators independently label the components in the design artifacts (e.g., the component belongs to \textit{not-icon} or \textit{$icon_x$}, as one artifact may contains multiple icons, we apply $x$ to annotate icon separately).
Each design artifact was labeled by 3 different annotators and the ground truth was generated until an agreement was reached.
All annotators have previous experience designing and implementing icons.
In total, annotators labeled 1,012 real-world design artifacts with 9,883 icons.

\subsubsection{Baselines \& Metrics:}
Since the number of icons in the design artifact is unknown, we set up two widely-used clustering algorithms that do not need to pre-set the number of clusters in advance as baselines, Mean-Shift Clustering~\cite{cheng1995mean} and Density-Based Spatial Clustering of Applications with Noise (DBSCAN)~\cite{ester1996density}.
To further gauge the advantages of our heuristic icon composition algorithm, we compare the icon produced by our correlation metric to each metric individually.

\textit{Mean-Shift:}
Mean-shift is a centroid-based clustering algorithm to detect the close components in the design artifact. 
It works by updating the candidates for the center points as the mean of the points within the sliding window. 

\textit{DBSCAN:}
It is an improvement over the Mean-Shift clustering as it involves a transitivity based chaining-approach to determine whether components are located in a particular cluster, in order to separate clusters of noise and outliers.

\textit{ATTR only:} It composes the icon based on the component's attribute in the artifacts, such as the curve components are merged together as an icon.

\textit{HRCHY only:} As designers may manage all the resources of the icon under a layout folder, we rely on this hierarchical structure information to assemble the icon.

\textit{IOU only:} The rendering of components in the icon are likely intersecting and overlapping to each other, therefore we compose the icon based on measuring the IOU of rendering.

To evaluate the performance, we set up three evaluation metrics, e.g., precision, recall, F1-score.
Precision is the proportion of icons that are correctly composed among all icons in the design artifacts.
Recall is the proportion of icons that are correctly composed among all composed components (e.g., \textit{$icon_x$} and \textit{not-icon}).
F1-score is the harmonic mean of precision and recall, which combine both of the two metrics above.

\begin{table}[t!]
    \centering
    \begin{tabular}{l|cc|ccc|c}
    \toprule
    \bf{Method} & Mean-Shift & DBSCAN & ATTR only & HRCHY only & IOU only & Our \\
    \hline
    \bf{Precision} & 40.16\% & 55.54\% & 27.10\% & 3.23\% & 45.40\% & \bf{76.50\%} \\
    \hline
    \bf{Recall} & 51.91\% & 73.34\% & 35.20\% & 6.25\% & 62.51\% & \bf{81.25\%} \\
    \hline
    \bf{F1-score} & 45.28\% & 63.21\% & 30.62\% & 4.26\% & 52.59\% & \bf{78.80\%} \\
    \bottomrule
    \end{tabular}
    \caption{Performance comparison for icon composition.}
    \label{tab:compositionResult}
\end{table}

\begin{figure}
	\centering
	\subfigure[Compared to \textit{ATTR only}, our algorithm can compose icon with different attributes (bitmap and path).]{
		\includegraphics[width = 0.3\textwidth]{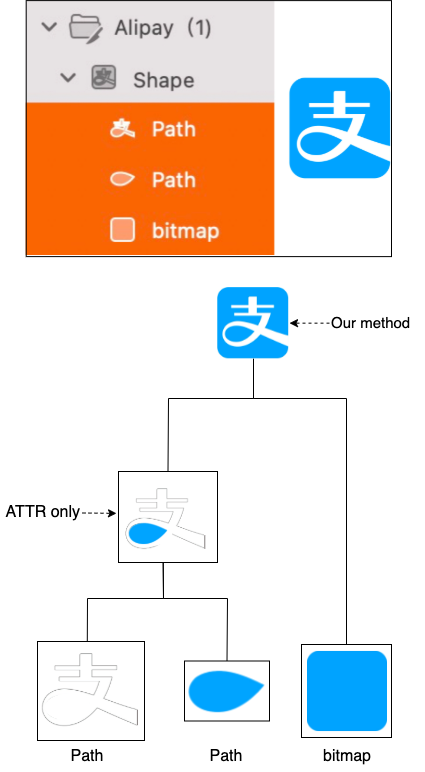}
		\label{fig:example1}}
	\hfill
	\subfigure[Compared to \textit{HRCHY only}, our algorithm can compose icon with different hierarchies ("Oval" not in the layout folder).]{
		\includegraphics[width = 0.3\textwidth]{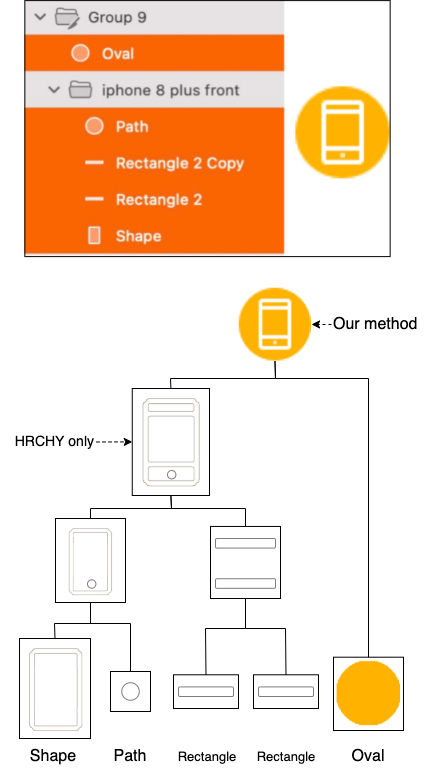}
		\label{fig:example2}}	
	\hfill
	\subfigure[Compared to \textit{IOU only}, our algorithm can compose icon with no intersection (shapes are widely spaced).]{
		\includegraphics[width = 0.3\textwidth]{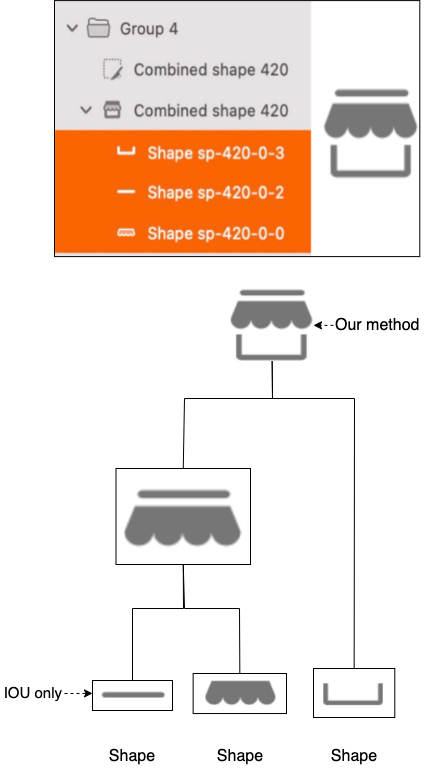}
		\label{fig:example3}}	
	\caption{The results of our icon composition algorithm. The top represents the design artifact, and the bottom represents the clustering steps.}
	\label{fig:iconresult}
\end{figure}

\subsubsection{Results:}
Table~\ref{tab:compositionResult} shows the performance of all methods.
The centroid-based clustering algorithm can only achieve 45.28\% and 63.21\% F1-score, for \textit{Mean-Shift} and \textit{DBSCAN} respectively.
The issues with these baselines are that they are designed for large and dense data, such as natural disasters, stations, etc.
However, different from those data, the number of components in design artifacts is relatively small.
The performance of our algorithm is much better than that of other correlation baselines, e.g., 31.1\%, 18.74\%, 26.21 boost in precision, recall, and F1-score compared with the best correlation baseline (\textit{IOU only}).
Relying only on hierarchical structure information (\textit{HRCHY only}) yields poor performance in all metrics, 3.23\%, 6.25\%, and 4.26\% respectively compared to the ground truth.
This is because that many designers follow their own design styles which highly differ from each other, and some of their design miss the rigid layout information. 
\textit{ATTR only} yields an average F1-score of 30.62\%, which is better than the F1-score of \textit{HRCHY only}, indicating that the attribute of components provides better information for icon composition as designers tend to use consistent attributes for icon designs.
\textit{IOU only} yields the best performance among all baselines with the average F1-score to 52.59\%, indicating that the icons are normally composed by several intersecting components.
Fig~\ref{fig:iconresult} shows some results of icon composition by our heuristic algorithm. 
It shows that our algorithm can help cluster the components with different attributes (e.g., bitmaps and path for the first icon), different hierarchies (e.g,  "Oval" component is not in the layout for the second icon), and no intersection (e.g., the components of icon have spacing for the third icon).


To identify the common causes of errors, we further manually check the wrong composition cases in the test dataset. 
According to our observation, we found two main reasons.
First, the components of icons are widely spaced, such as, three separated waves for "wifi", three horizontal lines for "menu", several lines from low to high for "signal", etc.
In addition, designers may depict some effects around the icon, such as, the ripple effect on a hand for "tap", the shiny effect on a bulb for "light", etc.
The wide space between components decreases the correlation value of IOU, resulting in poor clustering.
Second, some designers place icon and text in the same hierarchy to depict an image button, for example, a folder consists of icon components, text components, button border, etc., causing our algorithm to aggregate an image button, rather than an icon.

\subsection{Icon Prediction}
\subsubsection{Dataset:}
We leveraged the categorization during the creation of the semantic vocabulary (in Table~\ref{tab:categorization}), and corresponding icons and attached labels as the training data.
The foundation of the deep learning model is the big data, so we only selected categories with frequency larger than 300 for training the model.
Therefore, there are 100 categories left with the number of icons ranging from 311 to 589. 
Given all the data assigned to each label, we randomly split these 41k icons into train/validation/test dataset with a ratio of 8:1:1 (33K:4K:4k).
To avoid the bias of of “seen samples” across training, validation and testing, we performed 5-fold cross-validation.

\subsubsection{Baselines:}
We set up several basic machine-learning baselines including the feature extraction (e.g., color histogram~\cite{wang2010robust}, scale-invariant feature transform~\cite{lowe1999object}) with machine-learning classifiers (e.g., decision tree~\cite{quinlan1983learning}, SVM~\cite{cortes1995support}).
Apart from these conventional machine learning based baselines, we also set up several derivations of state-of-the-art deep learning models as baselines to test the importance of different inputs of our approach including backbones (ResNet~\cite{he2016deep}, VGG~\cite{simonyan2014very}, MobileNet~\cite{sandler2018mobilenetv2}), different input channels (RGB, RGBA).  
The training and testing configurations for these baselines were the same.

\begin{table}[t!]
    \centering
    \footnotesize
    \begin{tabular}{l|cccc|cccc}
    \toprule
    \bf{Method} & \specialcell{\bf{Histo} \\ \bf{+SVM}} & \specialcell{\bf{Histo} \\ \bf{+DT}} & \specialcell{\bf{SIFT} \\ \bf{+SVM}} & \specialcell{\bf{SIFT} \\ \bf{+DT}} & \bf{ResNet-50} & \bf{VGG-16} & \specialcell{\bf{MobileNetV2} \\ \bf{(RGBA)}} & \specialcell{\bf{MobileNetV2} \\ \bf{(RGB)}}\\
    \midrule
    \bf{Accuracy} & 0.5657 & 0.3267 & 0.5806 & 0.4686 & \bf{0.8839} & 0.8764 & 0.8348 & 0.8772 \\
    \hline
    \bf{Time (ms)} & 0.103 & 0.152 & 1.702 & 1.941 & 26.535 & 27.282 & 17.567 & \bf{17.485} \\
    \bottomrule
    \end{tabular}
    \caption{Label classification accuracy and time estimation in different methods.}
    \label{tab:predictresultTable}
\end{table}

\subsubsection{Results:}
\label{prediction_result}
As we trained a classifier to predict label for icon, we adopted the accuracy as the evaluation metric for the model, illustrated in Table~\ref{tab:predictresultTable}.
The traditional machine learning method based on the human-crafted features can only achieve about 0.6 average accuracy.
Deep learning models perform much better than the best old fashioned methods, i.e., with the 0.3033, 0.29712, 0.2958 increase for ResNet-50, VGG-16, and MobileNetV2 respectively.
Although ResNet model performs the best in icon classification task, it requires relatively long time for prediction (26.535ms per icon) which strongly violates the performance of UI rendering (16ms) as we aim to deploy into an online platform (\textit{Imgcook}).
In contrast, our model MobileNet is nearly as accurate as ResNet with a performance lag of 0.67\%, while being 34.1\% faster.
And also, we find that the increase of a fourth alpha channel (RGBA) decreases the accuracy from 0.8772 to 0.8348, due to two main reasons.
First, the result shows that the model with RGB input has a loss value of 0.7844 at epoch 200, which is better than the model with RGBA (0.9231). This is because the supplemented channel greatly increases the parameters of the model, which leads to a decline in the ability of gradient training at the same epochs.
Second, based on the principle of optics, the fourth alpha channel does not reflect the morphological characteristics of the image. It is used to reduce information of other three channels by adjusting their color/degree, causing less information captured through training process.

\subsection{User Study}

\begin{table*}[t!]
\caption{Examples of development for icons in \textbf{E}xperimental and \textbf{C}ontrol groups.}
\centering
\begin{tabular}{p{0.3\linewidth}|p{0.05\linewidth}p{0.5\linewidth}}
    \toprule
    \multirow{4}{*}{\begin{minipage}[t]{2em}\centering\includegraphics[width=1.75in]{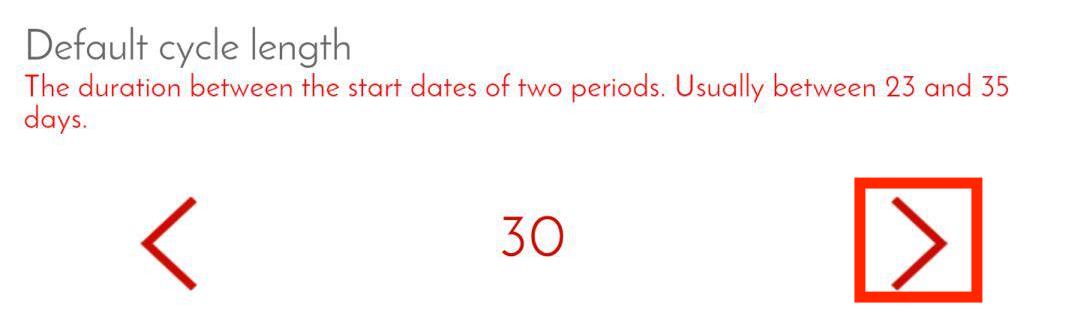}\end{minipage}} & \bf{E} & \small{<i class="icon-left red"></i>} \\ \cmidrule{2-3}
    & \bf{C1} & \small{<img src="8E431911-61BB-4A19-8C01.svg"/>} \\
    \cmidrule{2-3}
    & \bf{C2} & \small{<img src="next-icon-design.svg" alt="next" width="100\%"/>} \\
    \cmidrule{2-3}
    & \bf{C3} & \small{<div style="background-image: url('8E431911-61BB-4A19-8C01.svg');"></div>} \\
    \midrule
    \multirow{4}{*}{\begin{minipage}[t]{2em}\centering\includegraphics[width=1.65in]{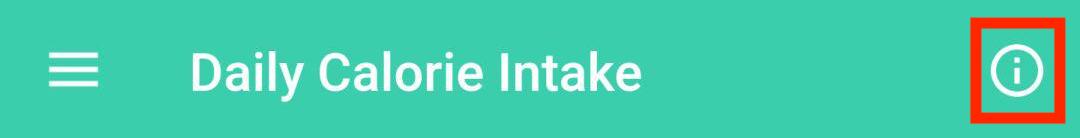}\end{minipage}} & \bf{E} & \small{<i class="icon-information white"></i>} \\ \cmidrule{2-3}
    & \bf{C1} & \small{\begin{tabular}[c]{@{}l@{}}\textless{}svg class="icon" aria-hidden="true"\textgreater\\   \hspace{0.5cm} \textless{}use xlink:url="icon-information"\textgreater{}\textless{}/use\textgreater\\ \textless{}/svg\textgreater{}\end{tabular}} \\
    \cmidrule{2-3}
    & \bf{C2} & \small{<a style="background-image: url('q\&a.svg'); width: 100\%; height:100\%;" class="help-icon"></a>} \\
    \cmidrule{2-3}
    & \bf{C3} & \small{<a class="svg-icon-info" href="\#"><SvgIcon name="white"></a>} \\
    \bottomrule
\end{tabular}
\label{tab:icon_development}
\end{table*}


\subsubsection{Procedures for User Study}
10 developers (generators), all proficient in UI development and have at least 1-year experience, were recruited for this study.
We randomly selected five icons from the real-life UI designs and asked each generator to develop them.
To guarantee the generators can objectively develop the icons, we asked whether they have prior knowledge on the icons (such as development experience, design experience, etc.).
The time of the development were recorded.
To be fair, generators did not know we were recording the time as the time pressure may affect their development (in quality, speed, etc.)~\cite{austin2001effects,mantyla2014time}.
We set the manual development as the control group.
Then, we also asked them to develop five other icons with the help of our tool which not only automatically convert the image to font, but also provide the description (predicted label and color).
We called this the experimental group.
The detailed developments of two groups for icons are shown in Table~\ref{tab:icon_development}. 

We then recruited another 10 developers (evaluators), and each of them was assigned the developments from two control groups and one experimental group.
Note that they did not know which one if from the experimental or control group, and for each icon, we randomly shuffled the order of candidates to avoid potential bias.
Given each development, they individually marked it as readable or not in five-point likert scale (1:not readable at all and 5:strongly readable).
To evaluate the performance of usability, we also asked evaluators to rate how likely they would like to use the development in practice (Acceptable).
The measurement is also in five-point scale.


\begin{figure}
    \centering
    \begin{minipage}{.65\textwidth}
        \centering
        \includegraphics[height=1.7cm]{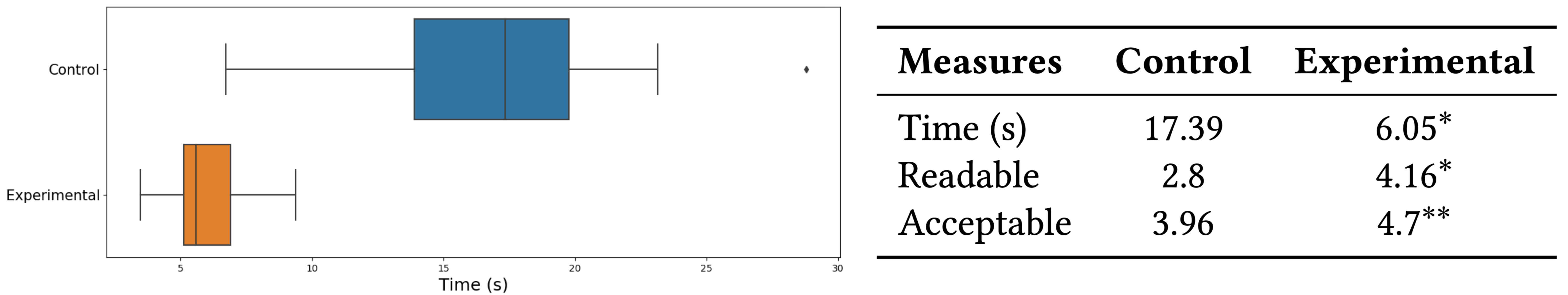}
        \caption{The comparison of Experimental and Control groups. $^*$denotes $p<$ 0.01, $^{**}$denotes $p<$ 0.05.}
        \label{fig:time}
    \end{minipage}%
    \hfill
    \begin{minipage}{.28\textwidth}
        \centering
        \includegraphics[height=1.8cm]{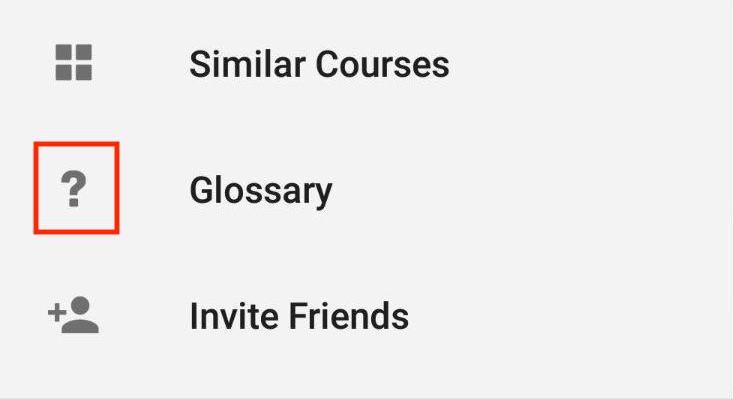}
        \caption{The icon description varies by context.}
        \label{fig:serendipitous}
    \end{minipage}
\end{figure}


\subsubsection{Results:}
\label{user_study_result}
Box plot in Fig~\ref{fig:time} shows that the time spent on the development of icons in the experimental group is much shorter than that in the control group (with an average of 6.05s versus 17.39s, saving 65.2\% of time).
That is the biggest strength of our tool i.e., developers can quickly develop an icon by providing descriptions and a font pattern.
On average, the overall readability ratings for the experiment group is 4.16, which are significantly higher (48.5\%) than the control group (2.8) in Fig~\ref{fig:time}. 
Most developers admit that our tool can provide more acceptable results for them.
In other words, 94\% (4.7/5.0) of developers hope to develop the icons with the help of our tool in their real development environment compared to 3.96 in the control group.
To understand the significance of the differences, we carry out the Mann-Whitney U test~\cite{fay2010wilcoxon} (specifically designed for small samples) on the readability and acceptability ratings between the experiment and the control group respectively. 
The test results suggest that our tool does significantly outperform the baseline in term of these metrics with $p < 0.01$ or $p < 0.05$.

For some icons, the developer gives very low acceptability score to the labels. 
According to our observation and further survey, we summarise two reasons accounting for those bad cases. 
(1) Albeit the good performance of our model, we still make wrong predictions for serendipitous icons. Based on the context of icon, the same icon can have different meaning.
For example, the icon in Fig~\ref{fig:serendipitous} represents the meaning of "information" in the common case, but consider the text on the right, the meaning of the icon should be "glossary"/"dictionary". 
(2) Developers admit the usefulness of converting images to font for providing faster rendering speed.
However, they also point out the limitation of replacing image with font.
Font is not fully compatible in all browsers and devices. One developer mentioned that they need to make sure that the development works on old devices, in which they usually need to give up latest efficient methods, such as iconfont. 

\subsection{Industrial Usage}
We cooperate our tool with the \textit{Imgcook} platform\cite{imgcook} developed by Alibaba, an intelligent tool to automatically generate front-end codes from UI design files.
\textit{Imgcook} has attracted a lot of attention in the community which has a large user base (15k) and generates over 40k UIs.
\tool is integrated with the internal automated code generation process and is triggered whenever the design files contain an icon. 


In order to evaluate the usability of our tool, we set up a code review metric for measuring the code modification for icons. 
Note that the code modification contains multiple contents, such as text, button, etc, we only measure the modification if the object is icon to reduce the potential bias.
We adopt a case-insensitive BLEU (BiLingual Evaluation Understudy)~\cite{papineni2002bleu} as the metric to evaluate the preservation of code. BLEU is an automatic evaluation metric widely used in code difference studies. It calculates the similarity of machine-generated code and human-modified reference code (i.e., ground truth) as
\(BLEU = BP * exp(\sum_{N}^{n=1} w_{n}logp_{n})\)
where $p_{n}$ denotes the precision, i.e., the ratio of length n token sequences generated by our method are also present in the ground-truth; $w_{n}$ denotes the weight of different length of n-gram summing to one; $BP$ is 1 if c > r, otherwise $e^{(1-r/c)}$ where c is the length of machine-generated sequences and r is the length of ground-truth.
A high BLEU score indicates less modification in the code review.


We run the experiment in \textit{Imgcook} with 6,844 icons in 2,031 UIs from August 20, 2020 to September 20, 2020.
Among all the testing UI developments, the generated code for icon reaches 84\% BLEU score, which means that most of the code is used directly without any modification. It demonstrates the high usability of \tool in practice. 
Based on the inspection results, we categorize the modification into four categories.
Two reasons are discussed in the Section~\ref{user_study_result}, in terms of wrong prediction and compatibility concern.
There are another two modifications mainly due to industrial practice. 
First, in order to maintain the consistency of company's coding experience, some developers modify to a prescribed naming/rendering method, for example, packing the icon of "icon-camera" to a <Icon-Camera> tag.
Second, UI dynamically changes in practice.
Once an element in the UI is changed, the attribute of icon may change, such as color and font size.


\noindent\fbox{
    \parbox{0.98\textwidth}{
        Overall, our method achieves 78.8\% F1-score in the icon composition from design artifacts (RQ1), and 87.7\% accuracy in the label prediction for icons (RQ2).
        In the survey of 10 developers, we improve the efficiency of developing time and code readability by 65.2\% and 48.5\%, respectively (RQ3).
        The majority (4.7/5.0) of the interviewed developers acknowledges the usability of the generated code for icon by our method, and it is further confirmed in the practice of \textit{Imgcook} with 84\% BLEU score (RQ4).
    }
}

\section{Discussion}

\textbf{On developers:}

Due to the conceptual gap between designers and developers, there is a transition barrier from design artifact to implementation. Developers require extra effort to compose design components into icons for optimizing the network traffic and caching.
To bridge this transition gap, our work proposes a heuristic machine learning technique to cluster components in the design artifacts to automated agglomerate the scattered icon components into clusters, helping to ease the burden of manual search for developers. The clusters can be further applied to various downstream tasks such as icon implementation, UI testing~\cite{moran2018machine}, css clipping~\cite{larsen2018mastering}, etc.
Icon implementation is a challenging and time-consuming task, even for professional developers.
On the one hand, UI developers must enhance performance.
Poor development has an adverse effect on the performance of the site. 
The performance issues comprise a multitude of factors like rendering speed, reusability \& flow of the code, etc.
On the other hand, UI developers must write a clean, high quality code which can be easily understood and maintained.
Inspired by the high performance of font rendering, our work designs an automated method to convert icon to font using computer vision techniques to trace the edge of icon and using graphic algorithm to optimize the edge.
In addition, compared with the missing descriptions in the development or brainstorming suitable names which is limited to several developers in the physical world, our deep learning and computer vision techniques based method can quickly identify the label and the color of icon.
Our method once made accessible to developers, can very well help developers achieve efficient icon coding.


\textbf{On the generalization of our method:}
We report and analyze the performance of our CNN-based model for predicting icon labels in Section~\ref{prediction_result}. 
One limitation with our model is that we currently only test our model on 100 labels with enough corresponding icons.
With the cooperation with \textit{Imgcook} platform, the icons in the UI images are a gold resource.
First, the icons are relatively unique, otherwise, developers can reuse the online resources directly. These unique icons can significantly increase the amount of data, consequently improving the accuracy of our model.
Second, developers may modify the description to a serendipitous label which can augment the labels and generalize a broader range of icon descriptions.
Due to the time limit, we only collect a small amount of icons from \textit{Imgcook}.
However, we have seen some interesting icons that do not exists on online sharing platforms and they may improve the generalization of our method.

\textbf{Area of improvements:}
Currently, we only predict the label based on icon itself. As discussed in Section~\ref{user_study_result}, the meaning of icon varies in different context.
To address this problem, we can consider the entire UI, capturing all the related information to make the final prediction.
Developers praise the idea of adding descriptions to the code which is a tedious task for them.
They wonder whether our model can extend to other elements.
An developer hope us to support description for buttons as he finds many buttons do not have descriptive texts to explain its intention, resulting in a bad user experience.
We believe our model could help developers in this case as it will not be difficult to extend to other elements once we obtain enough data for the training.
Moreover, developers envision the high potential in being able to add icon size descriptions as one of the biggest strength of font is lossless scalability. To that end, we can measure the height of the icon and map it to the corresponding font size.

\section{Conclusion}

In this paper, we present a novel approach, \tool, that can provide developers with intelligent support to reduce the development time of icon design in the UI.
Our approach consists of two integral parts: a heuristic machine learning methods for icon composition, and a deep learning and computer vision method for icon implementation.
Our work possesses several distinctive advantages:
1) to bridge the conceptual gap between designers and developers in icon, we propose a heuristic-based clustering method to compose the scattered icon pieces in the design artifact into an icon.
2) we develop an automated image conversion method to turn an icon into a font in which improving icon rendering speed. 
3) to assist developers with better code accessibility, we adopt a deep learning model to automatically predict the descriptive label that convey the semantics of the icon.
4) base on the colorspace of the image, we detect the primary color of the icon to provide developers more knowledge on the icon.
Our method is incorporated into existing automated code generation platform to extend them beyond effective and descriptive coding.

\begin{acks}
We appreciate Yanfang Chang, Zixiao Zhao, and Jiawen Huang for designing and conducting a survey experiment, and all the participants from Imgcook team in Alibaba Group for taking surveys.
\end{acks}

\bibliographystyle{ACM-Reference-Format}
\bibliography{main}

\end{document}